\documentclass[%
twocolumns,
 reprint,%
 amssymb, amsmath,%
 aip,cha,%
groupedaddress 
]{revtex4-1}

\usepackage{docs}%
\usepackage{bm}%
\usepackage[colorlinks=true,linkcolor=blue, allcolors=blue]{hyperref}%
\usepackage{graphicx}

\expandafter\ifx\csname package@font\endcsname\relax\else
 \expandafter\expandafter
 \expandafter\usepackage
 \expandafter\expandafter
 \expandafter{\csname package@font\endcsname}%
\fi
\begin{document}

\title{Ferromagnet/Semiconductor/Ferromagnet Hybrid Trilayers grown using Solid-phase Epitaxy}%

\author{S. Gaucher}
\author{B. Jenichen}
\email{bernd.jenichen@pdi-berlin.de}
\author{J. Herfort}

\affiliation{Paul-Drude-Institut f\"{u}r Festk\"{o}rperelektronik Leibniz-Institut im Forschungsverbund Berlin e.V., Hausvogteiplatz 5-7, 10117 Berlin, Germany}%

\date{\today }

\begin{abstract}
The direct growth of semiconductors over metals by molecular beam epitaxy is a difficult task due to the large differences in crystallization energy between these types of materials. This aspect is problematic in the context of spintronics, where coherent spin-injection must proceed via ballistic transport through sharp interfacial Schottky barriers. We report the realization of single-crystalline ferromagnet/semiconductor/ferromagnet hybrid trilayers using solid-phase epitaxy, with combinations of Fe$_3$Si, Co$_2$FeSi, and Ge. The slow annealing of amorphous Ge over Fe$_3$Si results in a crystalline film identified as FeGe$_2$. When the annealing is performed over Co$_2$FeSi, reflected high-energy electron diffraction and X-ray diffraction indicate the creation of a different crystalline Ge(Co,Fe,Si) compound, which also preserves growth orientation. It was possible to observe independent magnetization switching of the ferromagnetic layers in a Fe$_3$Si/FeGe$_2$/Co$_2$FeSi sample, thanks to the different coercive fields of the two metals and to the quality of the interfaces. This result is a step towards the implementation of vertical spin-selective transistor-like devices.
\end{abstract}

\maketitle

\section{Introduction}
The growth by molecular beam epitaxy (MBE) of ferromagnetic (FM) Heusler alloys and semiconductors (SC) heterostructures was found to be a successful approach to realize a number of applications in the field of spintronics.\cite{Palmstrom2003,Palmstrom2016} The possibility to create lattice-matched and sharp FM/SC interfaces is important to realize high spin injection efficiency, as intermixing and ordering effects are known to play a detrimental role on spin transport.\cite{Schultz2009, Adelmann2004, Zega2006, Demchenko2006}  Coherence is better preserved when spin injection proceeds via tunneling through a narrow Schottky barrier,\cite{Rashba2000, Hanbicki2002, Hanbicki2003} whose profile is also highly influential.\cite{Hu2011} Such studies are more easily performed in bilayer systems, where a single-crystalline FM film is grown over a SC substrate by MBE. Indeed, metals usually require lower crystallization temperatures, which ensures that no undesirable mixing or byproducts occur at the SC/FM interface. Growing a SC over a metallic substrate is however more challenging, and can be a limiting factor for a number of envisaged  applications, such as vertical spin-selective devices made in the FM/SC/FM configuration.

Ge has been used as a candidate SC to grow on FM surfaces due to its relatively low crystallization temperature ($\sim$300$^{\circ}$C), which offers better chances to prevent  intermixing at the interface. The growth of Ge over ferromagnetic Heusler alloys was investigated by MBE,\cite{Yamada2012, Kawano2013Gebuffer, Jenichen2014Stacks, Hamaya2016} surfactant-mediated MBE\cite{Maafa2013, Kawano2016Growth} and more recently solid-phase epitaxy (SPE). \cite{Gaucher2017,Sakai2017} Using the latter approach, it was shown that the slow annealing of a thin amorphous Ge film over Fe$_3$Si can yield a compound identified as FeGe$_2$,\cite{Jenichen2018} over which it was possible to grow another single-crystalline Fe$_3$Si layer.

In this work, we extend the SPE technique to include Co$_2$FeSi within the trilayer stacks. Co$_2$FeSi and Fe$_3$Si have similar lattice parameters, and thus can both be grown by MBE in a lattice-matched way over GaAs and Ge. The two FM materials are known to have coercive fields differing by only a few Oe,\cite{Herfort2004,Hashimoto2005} which is an advantage to control the magnetization of the films independently. Co$_2$FeSi has the highest magnetic moment per unit cell and highest Curie temperature of all ferromagnetic Heusler alloys. It has also for long been expected to exhibit a half-metallic character. Although this property is being disputed, one can minimally expect the compound to offer a degree of spin polarization in the 40-60\% range.\cite{Makinistian2012}

Despite the relatively high reactivity of Co$_2$FeSi (compared to Fe$_3$Si), it was still possible to use the SPE approach to crystallize amorphous Ge films and achieve fully epitaxial trilayers. The realization of such heterostructures, including FM materials with different coercive fields, is a supplementary step towards the creation of new spin-selective devices that could be operated both as tunnel transistors and magnetic tunnel junctions.

\section{Experimental}
\begin{figure}[h]
\includegraphics[width=0.8\linewidth,angle=0, clip]{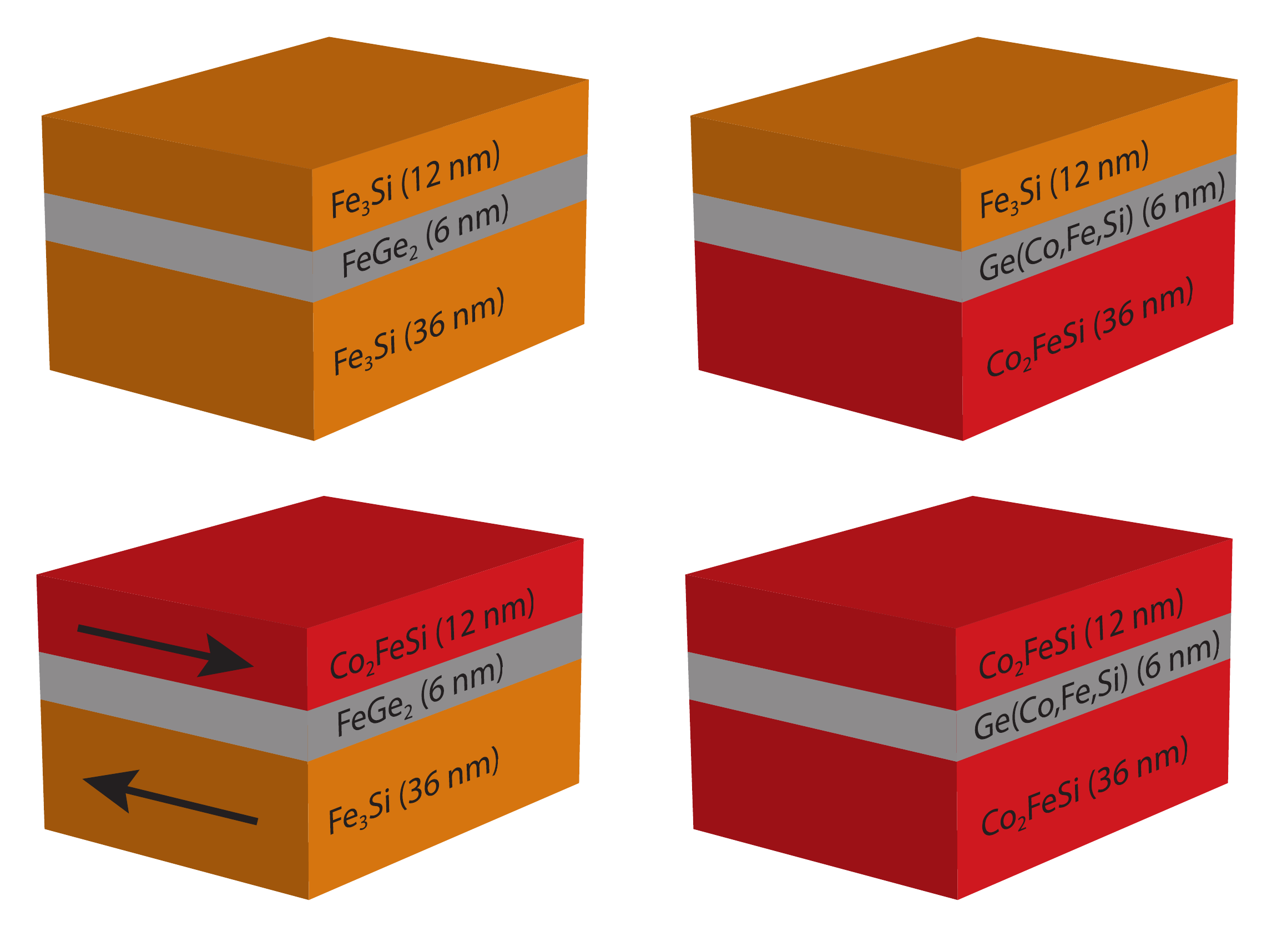}
\caption {Four trilayer stacking sequences containing a first 36~nm FM layer (Co$_2$FeSi or Fe$_3$Si), a 6~nm SC buffer layer (Ge crystallized by annealing), and a capping 12~nm FM layer (Co$_2$FeSi or Fe$_3$Si). An objective is to independently control the magnetization of each FM layer as illustrated on the bottom left stack. }
\label{stacks}
\end{figure}

Figure~\ref{stacks} shows four stacking sequences used in this study, where a Ge buffer layer is inserted between two FM films with thicknesses of 36 and 12~nm. The two FM layers are either Fe$_3$Si or Co$_2$FeSi, or a combination of both. The samples are grown using low-temperature MBE and SPE, building on methods that were described previously.\cite{Gaucher2017} GaAs(001) substrates are prepared with a 350~nm buffer layer, grown by MBE at 540$^{\circ}$C and As-terminated. The samples then are transferred under UHV into an As-free chamber, in which Fe$_3$Si and Co$_2$FeSi layers are grown by co-deposition from high-temperature effusion cells, both at 200$^{\circ}$C. On top of these first FM layers, the equivalent of 6~nm of crystalline Ge is deposited at 150$^{\circ}$C, which in fact results in an amorphous film. The samples are then annealed \textit{in-situ} by slowly increasing the temperature (5$^{\circ}$/min) up to 245$^{\circ}$C (over Co$_2$FeSi) or 260$^{\circ}$C (over Fe$_3$Si) for 10 minutes. During the annealing, the crystallization of the amorphous Ge layer is monitored by reflected high-energy electron diffraction (RHEED). The annealing temperatures were determined by observing the appearance of streaks in the RHEED pattern, which confirmed the obtention of a crystalline layer. Capping Fe$_3$Si or Co$_2$FeSi layers are then grown directly over the crystallized compounds, under the same conditions as the respective first layers.

The structure and quality of the trilayer stacks were evaluated by X-ray diffraction (XRD) using an X-Pert PRO MRD\textsuperscript{TM} system with CuK$\alpha_1$ radiation source having a wavelength $\lambda = 1.54056$~\AA. The FM layers are made with thicknesses in a ratio of 3:1 in order for their individual contributions to appear clearly in the XRD patterns (finite thickness oscillations with 1:3 beating pattern). The magnetization of the trilayers was measured in a superconducting quantum interference device (SQUID), using standard AC techniques. For that purpose, the samples were cut into pieces of approximately 3$\times$4~nm and cooled down to 10~K.

\section{Results}
\begin{figure}[h]
\includegraphics[width=0.85\linewidth,angle=0, clip]{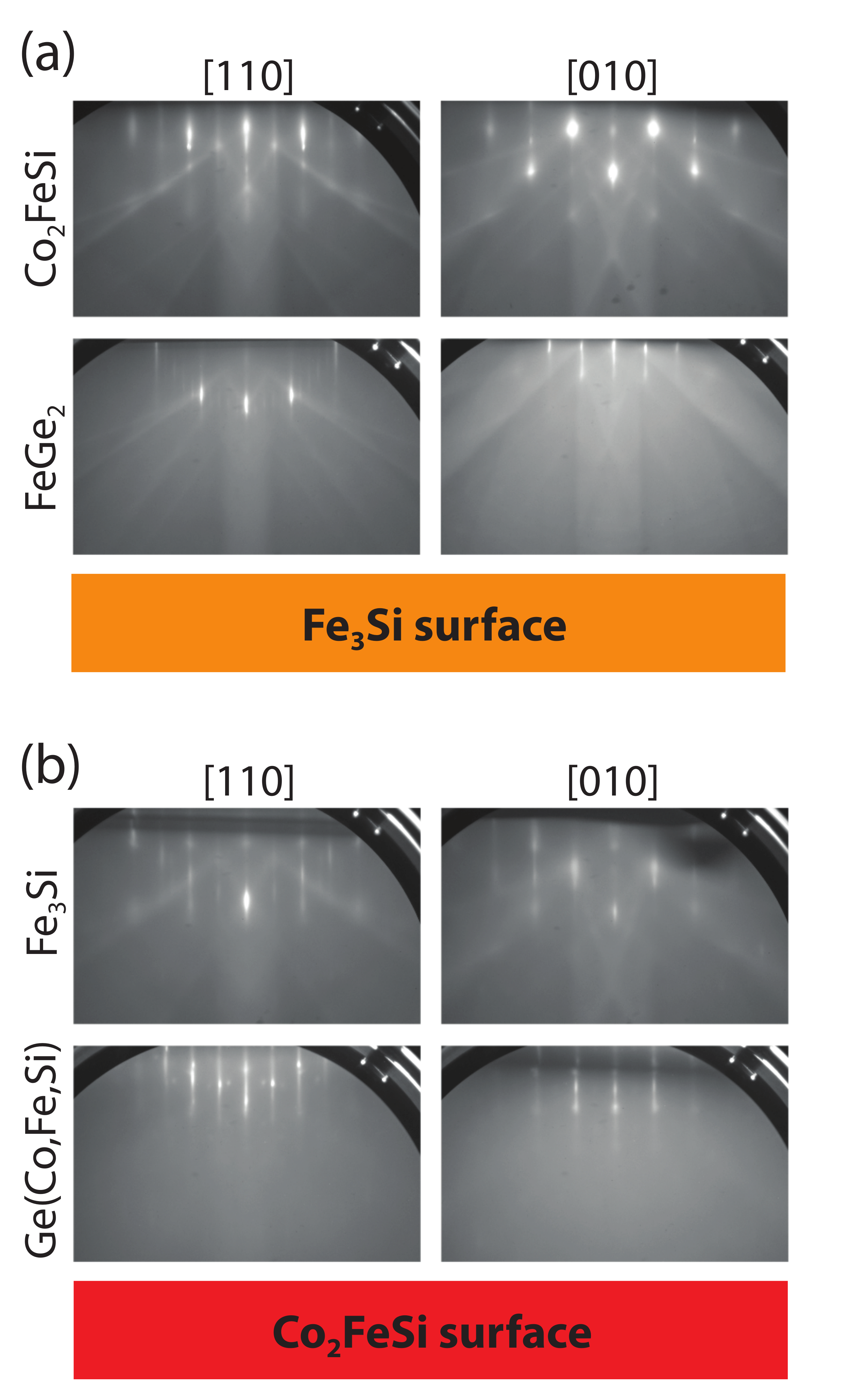}
\caption {(a) RHEED images taken during the growth of a Fe$_3$Si/FeGe$_2$/Co$_2$FeSi trilayer along the [110] and [010] azimuths. (b) Corresponding RHEED images taken during the growth of a Co$_2$FeSi/Ge(Co,Fe,Si)/Fe$_3$Si trilayer. The SPE of amorphous Ge yields different compounds depending on the nature of the underlying surface. It does not prevent the stacks from preserving the same crystal orientation, as seen from the alignment of the streaks of the capping FM films.}
\label{RHEED}
\end{figure}

All four hybrid trilayer configurations (illustrated in Figure~\ref{stacks}) have single-crystalline individual layers that preserve growth orientation. Figure~\ref{RHEED} shows a comparison of the RHEED patterns taken during the growth of two samples. In (a), the crystallization the amorphous Ge film is performed over a Fe$_3$Si surface. The resulting compound is a 2D allotrope of FeGe$_2$ with space group \textit{P4mm} (with a small amount of diffused Si atoms sitting on Ge sites). This material does not exist in a bulk form and results from the minimization of the elastic energy of the epitaxial film.\cite{Jenichen2018} There is a clear distinction between the patterns observed along the [110] and [010] directions. A relationship can thus be established with the orientation of the overgrown Co$_2$FeSi film, for which the streaks coincide for the same sample alignments. In (b), the annealing is done over a Co$_2$FeSi surface. Although the exact stoichiometry of the Ge(Co,Fe,Si) film was not determined, the RHEED pattern depicts a lattice-matched compound over which the pseudomorphic growth of Fe$_3$Si was possible. The absence of Kikuchi lines in the Ge(Co,Fe,Si) images is a sign that the surface is not as flat as the one obtained for FeGe$_2$, a feature that is also noticeable in the XRD curves of the trilayers.

\begin{figure}[h]
\includegraphics[width=0.95\linewidth,angle=0, clip]{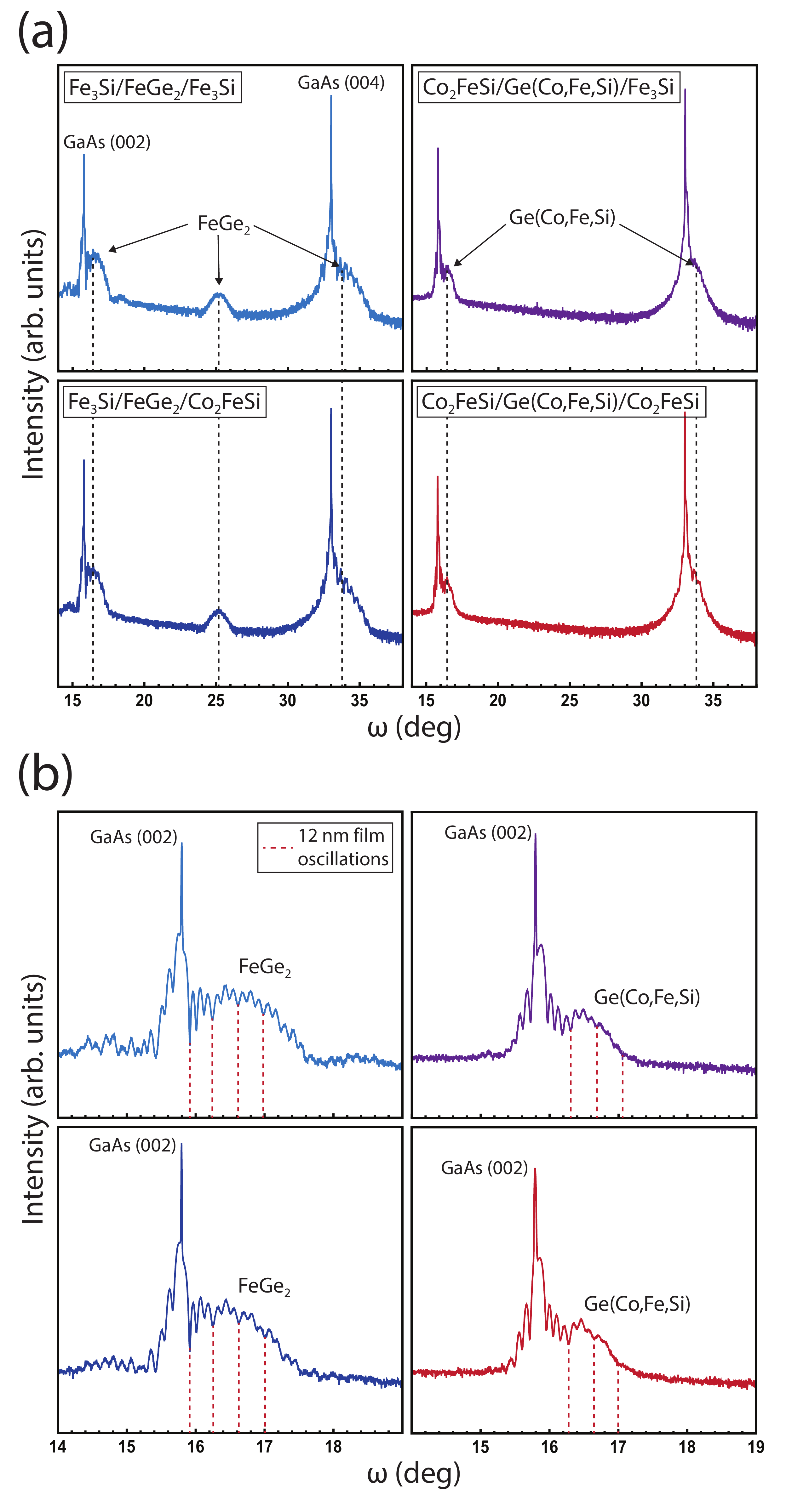}
\caption {XRD of four trilayer stacks, consisting of different combinations of Fe$_3$Si, Co$_2$FeSi and Ge compounds as indicated on each curves (see placement in Figure~\ref{stacks}). (a) Wide range diffraction curve showing a characteristic FeGe$_2$ peak about $\omega\approx$ 25$^{\circ}$ for samples with underlying Fe$_3$Si layer, as well as the crystalline Ge(Co,Fe,Si) peak over Co$_2$FeSi layers. (b) Close-up of the XRD curves about the GaAs(002) substrate peak, showing a beating pattern in the FM films oscillations. }
\label{xray}
\end{figure}

Figure~\ref{xray} shows the XRD curves of the four samples with different layer combinations. In (a), FeGe$_2$ is visible above Fe$_3$Si and corresponds to the rounded peaks at angles slightly higher than the GaAs(002) and (004) peaks ($\sim$16.5$^{\circ}$ and $\sim$ 34$^{\circ}$). The layered structure of the FeGe$_2$ is a superstructure with a period of two monolayers, which explains the satellite peak around $\omega\approx$~25$^{\circ}$.\cite{Jenichen2018} The Ge(Co,Fe,Si) forming over Co$_2$FeSi has peaks that coincide with the FeGe$_2$ ones. However, it does not have a layered structure, as can be asserted from the absence of a similar satellite peak. Panel (b) shows a close-up of the XRD curves near the GaAs(002) reflection peak. For all four trilayer stacks, it is possible to notice the contribution from both FM layers in the finite thickness oscillations. The 36:12~nm thickness ratio effectively generates a beating pattern between oscillations with periods in a 1:3 ratio. The clarity of these oscillations is better when the SPE is performed over Fe$_3$Si, which indicates that the stacks have better interface quality. The layered structure of the FeGe$_2$ is likely responsible for the sharpness of the interfaces. Thus, the SPE approach yields crystalline films over both Fe$_3$Si and Co$_2$FeSi surfaces, but the quality of the FM/FeGe$_2$ interfaces is better, which is an advantage in order to achieve clear and abrupt magnetization switching.

\begin{figure}[h!]
\includegraphics[width=0.95\linewidth,angle=0, clip]{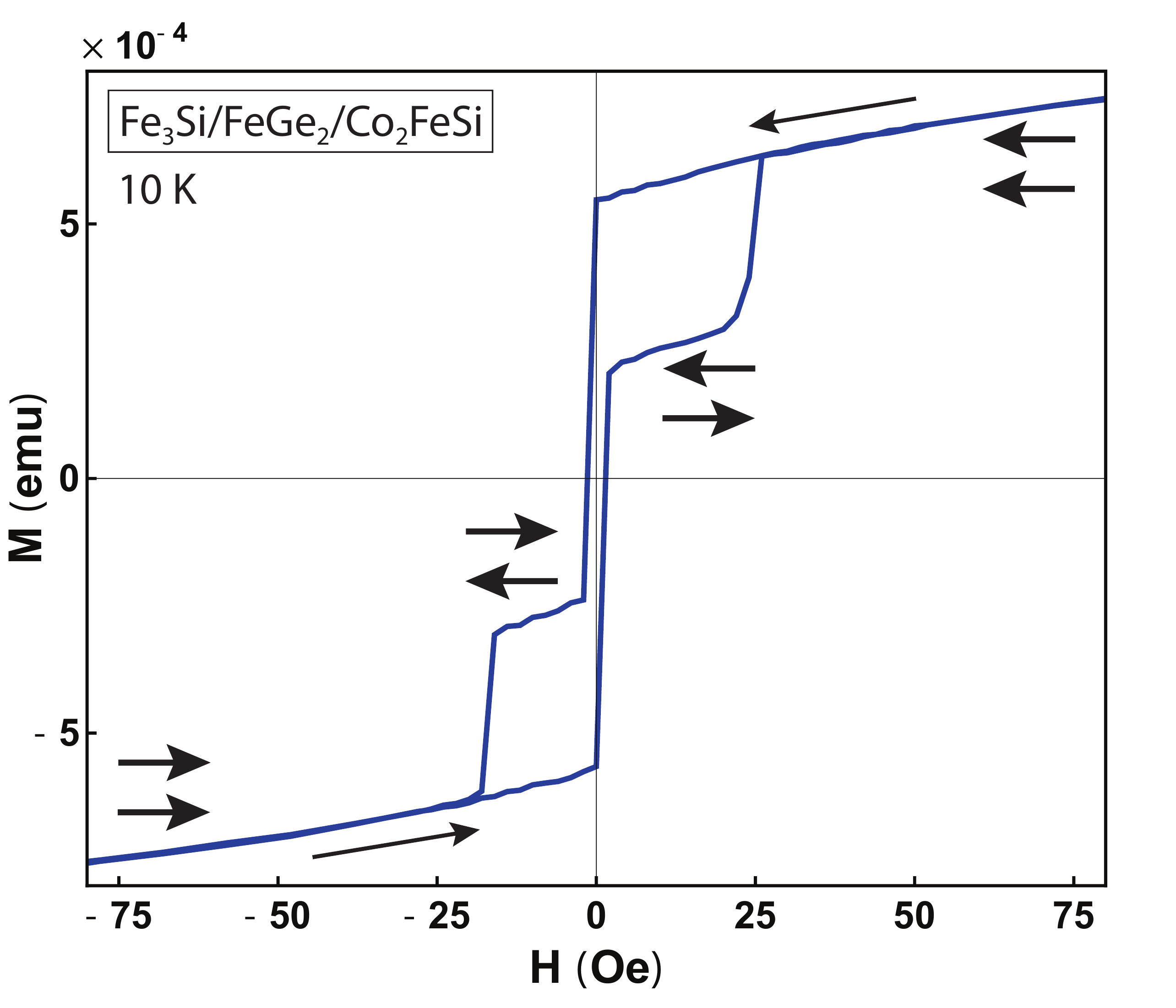}
\caption {SQUID magnetization hysteresis loop of a Fe$_3$Si/FeGe$_2$/Co$_2$FeSi sample cooled down to 10~K along the easy magnetization axis of the FM films. The curve shows clear intermediary plateaus  corresponding to antiparallel magnetization of the Fe$_3$Si and Co$_2$FeSi films. (The curve was recentered to correct for a known offset in the SQUID setup.)}
\label{squid}
\end{figure}

Figure~\ref{squid} shows the ideal behavior for the magnetization hysteresis of a trilayer. The sample used for this curve is the Fe$_3$Si/FeGe$_2$/Co$_2$FeSi stack, with external field applied along the easy magnetization axis of the FM films. At high positive field, both Fe$_3$Si and Co$_2$FeSi layers are magnetized along the same orientation. As the field is decreased, the magnetization of the thicker Fe$_3$Si layer flips first, resulting in a plateau. The Co$_2$FeSi layer then flips at slightly higher negative field, due to the higher coercivity of the material. Thinner films also tend to have higher coercive fields, which in this case accentuates the effect. A similar sequential switching is observed along the other direction of the hysteresis loop. The 6~nm FeGe$_2$ is also ferromagnetic at low temperatures. However, its contribution to the total magnetization is negligible and therefore cannot be observed on this curve.

It was not possible to observe clear independent magnetic switching of the FM layers using the samples having another stacking order. Single-crystalline Fe$_3$Si films are known to have small coercive fields and abrupt magnetization reversal (within 1~Oe). Despite having ideal interface quality, the different thickness of the FM films in the Fe$_3$Si/FeGe$_2$/Fe$_3$Si sample did not translate into a coercivity difference sufficient to produce  plateaus of antiparallel magnetization. The samples with underlying Co$_2$FeSi have rougher interfaces, which resulted in ill-defined features around the magnetization reversal which could not be clearly attributed to a specific FM layer. The combination of Fe$_3$Si and Co$_2$FeSi with FeGe$_2$ buffer layer is therefore the best all four stacking configurations.

\section{Conclusion}
The growth of FM/SC/FM trilayer stacks was investigated by a combination of low temperature MBE and SPE. The crystallization of amorphous Ge  was successful on both Fe$_3$Si and Co$_2$FeSi surfaces. On Fe$_3$Si, a layered allotrope of FeGe$_2$ arises spontaneously, while another Ge(Co,Fe,Si) compound forms over Co$_2$FeSi. In both cases, capping Fe$_3$Si or Co$_2$FeSi layers could be added by MBE while preserving the same growth orientation, as confirmed by the \textit{in-situ} RHEED patterns. XRD revealed the contribution from all three layers, independently of the order in which the layers were grown. However, higher interfacial quality is achieved when SPE is performed over Fe$_3$Si. Using a Fe$_3$Si/FeGe$_2$/Co$_2$FeSi sample, it was possible to observe clear independent magnetization switching of the FM films. The different thickness and coercive fields of the materials, the quality of the crystalline films, as well as the sharpness of the FM/FeGe$_2$ interfaces are responsible for the appearance of the effect in this specific sample. The successful realization of such hybrid FM/SC/FM trilayers is a step towards the realization of vertical spintronics devices.

\section{Acknowledgments}
The authors thank C.~Herrmann and H.-P.~Sch\"{o}nherr for their support during the growth of the samples.


%

\end{document}